\documentstyle[11pt,epsf]{article}

\newcommand{\beq} {\begin{equation}}
\newcommand{\eeq} {\end{equation}}
\newcommand{\what}[1]{\widehat #1}

\newcommand{\lh} {l_h}
\newcommand{\jh} {j_h}
\newcommand{\mh} {m_h}

\newcommand{\lp} {l_p}
\newcommand{\jp} {j_p}
\newcommand{\emp} {m_p}

\newcommand{\bsigma}{\mbox{\boldmath $\sigma$}}

\newcommand{\bkey}{{\bf k}}
\newcommand{\bpi}{{\bf p}}
\newcommand{\bqu}{{\bf q}}
\newcommand{\br}{{\bf r}}
\newcommand{\half}{\frac{1}{2}}

\begin{document}

\begin{titlepage}
\thispagestyle{empty}
\begin{center}
{\Large \bf Short-range correlations in semi-exclusive electron
            scattering experiments}

\vspace{1.5cm}
{\large S.R. Mokhtar$^{\,1,2}$, M. Anguiano$ ^{\,2,3}$, 
G. Co'$ ^{\,\,2,3}$ and
A.M. Lallena$^{\,4}$} \\ 
\vspace{1.cm}
{$^{1)}$  Department of Physics, University of Assiut, \\
        Assiut, Egypt} \\
\vspace{.5cm}
{$^{2)}$ Dipartimento di Fisica,  Universit\`a di Lecce, \\
I-73100 Lecce, Italy} \\
\vspace{.5cm}  
{$^{3)}$  Istituto Nazionale di Fisica Nucleare  sez. di Lecce, 
\\ I-73100 Lecce, Italy} \\
\vspace{.5cm}
{$^{4)}$ Departamento de F\'{\i}sica
Moderna, Universidad de Granada, \\
E-18071 Granada, Spain}
\end{center}
\vskip 1.5 cm 
\begin{abstract}
One-nucleon emission electron scattering experiments are studied with
a model that considers short--range correlations up to the first order 
in the number of correlation lines.
The proper normalization of the many-body wave functions requires the
evaluation of two- and three-point diagrams, the last ones
usually neglected in the literature. 
When all these diagrams are included the
effects of the short-range correlations are rather small.
The results of our calculations are compared with experimental data 
taken on $^{16}$O.
\end{abstract} 
\vskip 1.cm
PACS number(s): 21.10.Ft, 21.60.-n
\end{titlepage}

\section{Introduction}
One-nucleon knock out experiments induced by electromagnetic probes
have been, and still are, an important tool to study the structure of
the atomic nucleus \cite{fru84,bof96}.  The main use of these
experiments has been the investigation of the ground state single
particle structure of nuclei \cite{wit90}.  For this reason, they have
been done, in the great majority of cases, in the quasi-elastic
region, in order to minimize the effects of the collective nuclear
excitations.  Nucleon emission in the quasi-elastic region is
dominated by the direct knock out of the detected nucleon while the
other nucleons behave like spectators. Many-body effects can be
treated as corrections to this basic process.

These (e,e'p) experiments have been analyzed in terms of mean field
models with great success. The models describe the target nucleus
ground state by using a purely real mean field, and the nuclear
excited states in terms of one-particle one-hole excitation with the
particle moving in a complex optical potential.  The optical potential
is usually taken from a fit to proton elastic scattering data off the
A-1 nucleus.  This approach reproduces rather well the behavior of
the cross section in terms of the so-called missing momentum.  What
remains, however, is the problem of the overestimation of its magnitude.
The quenching factor necessary to reproduce the data has been related
to the spectroscopic factor which nowadays has achieved the same
status as a true observable \cite{wag82}.

The increased accuracy of the data as well as the possibility of
performing more elaborate numerical calculations, has stressed the
need for better descriptions of (e,e'p) data. The source of the
spectroscopic factor has been investigated following various
hypotheses, from relativistic effects \cite{udi93} to in medium
modifications of the nucleon properties \cite{ste86}.  At the moment,
one of the most frequently considered hypotheses is the partial
occupation of the single particle levels produced by the short-range
correlations \cite{pan84}.  Various estimates of these occupation
probabilities have been done both in nuclear matter \cite{ram89,ben89}
and in finite nuclei \cite{ami97,fab00a}, but their values do not seem
to be compatible with the empirical spectroscopic factors.  On the
other hand, direct calculations of (e,e'p) cross sections with models
taking into account short-range correlations show noticeable effects
produced by these correlations \cite{giu94}-\cite{jan00}, in spite of
the fact that also in these cases spectroscopic factors are still
needed.  These calculations of the semi-inclusive cross section have
been done with simplified treatments of the correlations. In some
calculations only the lowest order cluster terms have been considered
\cite{giu94,ryc95,jan00}.  In other calculations \cite{gai00} the
overlap wave function, which in the mean field model corresponds to
the hole single particle wave function, has been extracted from
correlated one-body density distributions evaluated via more or less
sophisticated ground state calculations.

There are two weak points in these calculations.  A first one concerns
the inconsistency of the various inputs of the calculations. Single
particle wave functions and correlations are not linked by a unique
nuclear hamiltonian as they should be. The second weak point is
related to the fact that the proper normalization of the many-body
wave function is not preserved.

We have developed a model which takes into account all the diagrams
containing a single correlation line \cite{co00}.  Since the cluster
expansion conserves the sum rules order by order \cite{cla79}, the
normalization of the wave function is maintained.  This is immediately
seen in the application of the model to the description of the ground
state densities \cite{co95,ari97}.  The validity of our model
has been tested in nuclear matter by comparing our results with those
obtained in a calculation that considers all the cluster terms
\cite{ama98}.  The agreement between the two results is excellent,
showing that, for relatively simple operators, like the charge
operator, the first order approximation is reliable.

The inputs of our model, single particle wave functions and
correlations, are taken from Fermi Hypernetted Chain (FHNC)
calculations of the ground state of closed shell nuclei obtained by
minimizing the hamiltonian expectation value evaluated with realistic
and semi-realistic nucleon-nucleon potentials
\cite{ari96,fab00}.

In this paper we present the results of our investigation of (e,e'p)
reactions in $^{16}$O. We have analyzed the validity of various
approximations commonly adopted in the mean field description of
these processes.  It is not at all evident that approximations which
are within a certain theoretical framework should remain valid in a
different one. Moreover, if the aim is to disentangle subtle physical
effects, the uncertainty in the result due to some adopted assumption
can be comparable with the magnitude of the searched effect.

Summarizing in few words our findings we may say that short-range
correlation effects are extremely small and cannot explain the 
values of spectroscopic factors needed to reproduce the data.

\section{The cross section}
In this section we briefly recall the expressions of the coincidence
cross section used in our calculations. We work in natural units
($\hbar=c=1, e^2=1/137.04$) and employ the conventions of Bjorken and
Drell \cite{bjo64}. 
The initial and final electron four vectors are respectively 
$k \equiv (\epsilon,\bkey)$ and $k' \equiv (\epsilon',\bkey')$,
and we used the symbol $q \equiv (\omega,\bqu)= k - k'$ 
to indicate the four momentum transfer. The four momentum of the
emitted nucleon is $p \equiv (\epsilon_p,\bpi)$. The reference system,
shown in Fig. 1,
has been defined following the commonly used prescriptions.
The scattering plane is defined by the electron vectors $\bkey$ and
$\bkey'$, $\theta$ is the angle of the scattered electron, the
quantization axis is taken along the direction of $\bqu$, and the
angle $\theta_p$ and $\phi_p$ define the direction of the emitted nucleon.

\begin{figure}[hb]
\begin{center}
\vspace*{0.cm}
\hspace*{-0.0 cm}
\leavevmode
\epsfysize = 250pt
\epsfbox[70 200 500 650]{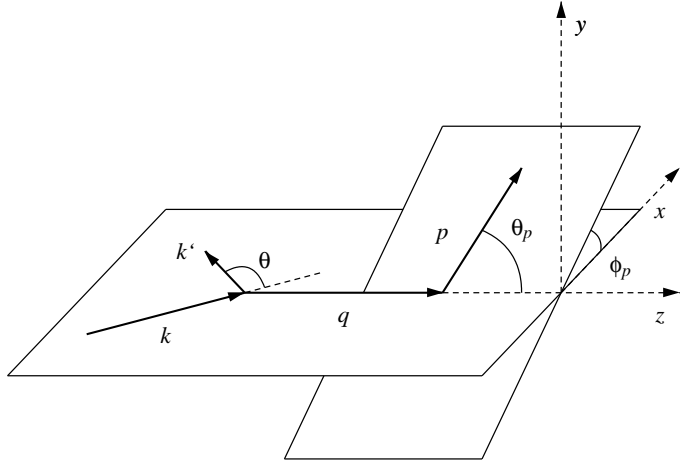}
\end{center}
\vspace*{-3cm}
\caption{\small Reference system used in our calculations.
}
\label{fig:axis}
\end{figure}

In our calculations the cross section has been derived by
adopting the commonly used assumptions \cite{for67}: 
the electron wave functions are plane wave solutions
of the Dirac equation, only one photon is exchanged between the electron
and the nucleus, all the terms depending upon the electron rest mass are
neglected. With these approximations we obtain:
\beq
\frac{{\rm d}^6 \sigma}
{{\rm d}\epsilon' {\rm d}\Omega_e {\rm d}\epsilon_p {\rm d}\Omega_p} =
\frac{\sigma_M K}{(2 \pi)^3}
\left(
  v_l w_l + v_t w_t + v_{tl} w_{tl} + v_{tt} w_{tt}   \right) \, ,
\label{cross}
\eeq
where we have indicated with $\sigma_M$ the Mott cross section:
\begin{equation}
\label{mott}
\sigma_M \, = \,
\left(\frac {e^2 \cos(\theta/2)}{2\epsilon_i \sin^2(\theta/2)}   
\right)^2
\end{equation}
Since we describe the emitted particle within the non relativistic
kinematics, we obtain $K=|{\bf {p}}| m_p$.
The factors $v$ come from the leptonic tensor and depend only from
kinematic variables:
\begin{eqnarray}
v_l &=& \left(\frac{q_\mu^2}{\bqu^2} \right)^2 \, ,\\
v_t &=& \tan^2 \frac{\theta}{2} - \half \frac{q_\mu^2}{\bqu^2} \, ,\\
v_{tl} &=& \frac {q_\mu^2}{\sqrt{2}\bqu^2} \,
 \left( \tan^2 \frac{\theta}{2} - 
\frac{q_\mu^2}{\bqu^2} \right)^\half \, ,\\
v_{tt} &=& \half \frac{q_\mu^2}{\bqu^2} \, .
\end{eqnarray}

The information about the nuclear structure is included in the $w$
factors. Because of the current conservation only three components of
the current four-vector are independent.
We choose the charge $\rho(\bqu)$ and the 
two transverse components in spherical coordinates:
\beq
J_{\pm}=\mp \frac{1}{\sqrt{2}} 
\left(J_x \pm i J_y  \right) \, .
\label{jpm}
\eeq
We can express the $w$ factors as \cite{co85,co87}:
\begin{eqnarray}
\label{wl}
w_l &=& \langle \Psi_i | \rho^\dagger (\bqu) | \Psi_f \rangle 
        \langle \Psi_f | \rho (\bqu) |\Psi_i \rangle  \, , \\ 
\label{wt}  
w_t &=& \langle \Psi_i | J_{-}^\dagger (\bqu) | \Psi_f \rangle 
        \langle \Psi_f | J_{-} (\bqu) |\Psi_i \rangle +
        \langle \Psi_i | J_{+}^\dagger (\bqu) | \Psi_f \rangle 
        \langle \Psi_f | J_{+} (\bqu) |\Psi_i \rangle \, ,\\  
\label{wtl}
w_{tl} &=& 2 Re  \left( \langle \Psi_i | \rho^\dagger (\bqu) | \Psi_f \rangle 
           \langle \Psi_f |  J_{-} (\bqu) |\Psi_i \rangle
        - \langle \Psi_i | \rho^\dagger (\bqu) | \Psi_f \rangle 
          \langle \Psi_i | J^\dagger_{+} (\bqu) | \Psi_f \rangle 
           \right) \, ,\\
\label{wtt}   
w_{tt} &=&  2 Re  \left(
           \langle \Psi_i | J^\dagger_{+} (\bqu) | \Psi_f \rangle 
           \langle \Psi_f |  J_{-} (\bqu) |\Psi_i \rangle
           \right) \, ,
\end{eqnarray}
where we have indicated with $|\Psi_i\rangle$
and $|\Psi_f\rangle$ initial and final states of the full hadronic
system.  In the previous equations a sum on
$\sigma$ and $m_h$ the projection on the z axis of the spin of
the emitted particle and of the angular momentum of the
the residual nucleus it is understood.

In this work we consider only one-body electromagnetic currents.
The charge operator is expressed as:
\begin{equation}
\label{charge1}
\rho({\bf r}) \, = \, 
\sum^A_{k=1} \frac{1+\tau^3_k}{2} \, 
\delta({\bf r}-{\bf r}_k) \, ,
\end{equation}
the convection current operator as:
\begin{equation}
\label{convection1}
J^{\rm C}({\bf r}) \, = \, 
\sum^A_{k=1} \frac{-i}{2M_k} \, \frac{1+\tau^3_k}{2} \,
\left[ \delta({\bf r}-{\bf r}_k)\nabla_k
- \nabla_k\delta({\bf r}-{\bf r}_k) \right] \, , 
\end{equation}
and the magnetization current operator as:
\begin{equation}
\label{magnetization1}
J^{\rm M}({\bf r}) \, = \, 
\sum^A_{k=1} \frac{1}{2M_k} \,
\left(\mu^{\rm P}\frac{1+\tau^3_k}{2} + 
\mu^{\rm N}\frac{1-\tau^3_k}{2} \right)
\nabla \times \delta({\bf r}-{\bf r}_k) \, \bsigma_k \, .
\end{equation}
In the previous equations $M_k$ indicates the rest mass of
$k$--th nucleon, 
$\mu^{\rm P}$ and $\mu^{\rm N}$ 
the anomalous magnetic moment of the proton and the
neutron respectively, $\bsigma_k$ the Pauli spin matrix of the
k--th nucleon and $\tau^3_k=1$ ($-1$) for protons (neutrons).
In our calculations the nucleonic internal structure has been
considered by folding the point-like
responses with the electromagnetic nucleon form factors of
Ref. \cite{hoe76}.

\section{The nuclear model}
The nuclear final state $|\Psi_f\rangle$ is described in the
asymptotic region as the product of the wave functions of the emitted
nucleon, $\phi(\bpi \cdot \br) \chi_\sigma$, with $\bpi$ its
momentum, and of the rest nucleus, $\Psi(\br_1 ...
\br_{A-1};\epsilon_h;\jh,\mh)$.  We have indicated with $\jh$ the
angular momentum of the rest nucleus and we have defined
$\epsilon_h=\epsilon_p-\omega$.  In a strict independent particle
model these quantities correspond to the energy and angular momentum
of the hole state.  We perform a multipole expansion of
$|\Psi_f\rangle$ to express it in terms of eigenstates of the total
angular momentum of the A nucleon system:
\begin{eqnarray}
\nonumber
|\Psi_f\rangle &=& \frac{4 \pi}{|{\bf p}|}
\sum_{\lp \mu_p} \sum_{\jp \emp} \sum_{JM,\Pi}
         i^{\lp} Y^*_{\lp \mu_p} (\what{\bpi}) 
         \langle \lp \mu_p \half \sigma | \jp \emp \rangle \\
\nonumber
&~&       \langle \jp \emp \jh \mh| J M \rangle
|\Psi; J M, \Pi; (\lp \jp \emp \epsilon_p,\lh \jh \mh \epsilon_h) \rangle \\ 
\nonumber
& \equiv & \frac{4 \pi}{|{\bf p}|}
\sum_{p} \sum_{JM,\Pi}
         i^{\lp} Y^*_{\lp \mu_p} (\what{\bpi}) 
         \langle \lp \mu_p \half \sigma | \jp \emp \rangle \\
&~&       \langle \jp \emp \jh \mh| J M \rangle \,\,
\frac { | J M, \Pi;p,h\rangle } 
{  \langle J M, \Pi;p,h| J M, \Pi;p,h\rangle^\half } \, .
\label{psif}
\end{eqnarray}
In the above equations the state 
$|\Psi; J M, \Pi; (\lp \jp \emp \epsilon_p, \lh \jh \mh \epsilon_h)
\rangle$ 
represents the excited state of the system with A nucleons with total
angular momentum $J$ projection $M$, parity $\Pi$, having a particle in
the continuum wave characterized by orbital and total angular momentum 
$\lp$ and $\jp$ with projection $\emp$ and energy $\epsilon_p$, and a
residual nucleus with hole quantum numbers $\lh$ $\jh$, $\mh$ and
$\epsilon_h$. We have indicated with $Y_{l \mu}$ the spherical
harmonics and with the symbol $\langle l_a m_a l_b m_b | J M \rangle$
the Clebsch-Gordan coefficients \cite{edm57}.
In the last expression, the nuclear excited state is given in
terms  of unnormalized many-body wave function $| J M, \Pi;p,h
\rangle$ 

In our model, the nuclear final state with total angular
momentum $J$ is described
as a superposition of all possible one-particle one-hole   
transitions allowed by the angular momentum selection rules. 
Within this approximation, the knowledge of the energy and of the
momentum of the emitted nucleon determines the quantum
numbers of the hole state of the residual nucleus.  
Then the summation on the hole states drops and
we sum only over the possible particle states in the
continuum.  We defined the sum over $p$ as a summation over the
$\lp$, $\mu_p$, $\jp$ and $m_p$ quantum numbers.

From the equations (\ref{wl})-(\ref{wtt}) defining the $w$
factors, 
one can see that the basic ingredients to be calculated are the
transition  
matrix elements from the ground state to the excited state of the
nuclear system induced by a one body operator. We write the
transition matrix elements as:
\begin{eqnarray}
\nonumber
\langle \Psi_f | O_{\eta}(\bqu) |\Psi_i \rangle &=& \frac{4 \pi}{|{\bf p}|}
\sum_{p} \sum_{J,M,\Pi} (-i)^{\lp} Y_{\lp \mu_p} (\what{\bpi}) 
         \langle \lp \mu_p \half \sigma | \jp \emp \rangle \langle \jp
         \emp \jh \mh| J M \rangle \\ 
\nonumber
&~&       \frac { \langle J M, \Pi;p,h|  O_{\eta}(\bqu) | 0 0, +1 \rangle} 
      { \langle J M, \Pi;p,h| J M, \Pi;p,h\rangle^\half 
        \langle 0 0, +1 | 0 0, +1 \rangle^\half  }
\\
&\equiv& \frac{4 \pi}{|{\bf p}|} \sum_{p} \sum_{J,M,\Pi} 
         {\cal  A}(p,h;J,M,\Pi) \,\, 
\xi_{J,M,\eta,\Pi}(\bqu;p,h) \, ,
\label{xi1}
\end{eqnarray}
where $| 0 0, +1 \rangle$
is the unnormalized ground state which, for the nuclei we shall
consider, has zero angular momentum and positive parity. We have included
all the geometric part of the expression in the symbol 
\beq
{\cal  A}(p,h;J,M,\Pi) =   (-i)^{\lp} Y_{\lp \mu_p} (\what{\bpi}) 
         \langle \lp \mu_p \half \sigma | \jp \emp \rangle \langle \jp
         \emp \jh \mh| J M \rangle \, , 
\label{xixixi2}
\eeq
and 
we have indicated with $O_{\eta}(\bqu)$ the generic one-body operator
inducing the transition. For $\eta=0$ this is the charge operator,
while for $\eta= \pm 1$ is given by the appropriate sum of convection
and magnetization currents, Eq. (\ref{jpm}).

In our nuclear model we consider the nuclear
states described as:
\begin{eqnarray}
\label{jasti}
| 0 0, +1\rangle &=& F \,\, |\Phi;0 0, +1\rangle \, , \\
\label{jastf}
| J M, \Pi;p,h\rangle &=& F \,\, |\Phi;J M, \Pi;p,h\rangle \, .
\end{eqnarray}
We have indicated with $|\Phi;0 0, +1\rangle$ the Slater determinant
describing the mean field wave function in a pure independent particle
model. This means that, taken a basis of single particle wave
functions, all the states below the Fermi surface are fully occupied 
and those above all completely empty. 
The state $|\Phi;J M, \Pi;p h\rangle$ indicates a Slater determinant where
the hole function $h$ has been substituted with the continuum
particle function $p$.
With respect to a pure independent particle model,
the novelty is the presence of the correlation function F. 
This function has, in principle, a very complicated
operatorial dependence, analogous to that of the
hamiltonian. Our calculations have been restricted to the
use of purely scalar (Jastrow) correlations. For this reason  
we immediately simplify the expressions formulating them only in terms
of this type of correlations. 
The adopted ansatz on the correlation is:
\beq
F(1,2,...A)=\prod^A_{i<j} f(r_{ij}) \, ,
\eeq
where $r_{ij}=|{\bf r}_i-{\bf r}_j|$ is the distance between the
positions of the particles $i$ and $j$.

Using the well-known cluster expansion techniques we rewrite the
transition matrix $\xi_{J,M,\eta,\Pi}(\bqu;p,h)$ in Eq. (\ref{xi1})
as:
\beq
\xi_{J,M,\eta,\Pi}(\bqu;p,h)  = 
\frac { \langle J M, \Pi;p,h|  O_{\eta} (\bqu) | 0 0, +1 \rangle} 
      { \langle 0 0, +1 | 0 0, +1 \rangle}
      \left[ \frac { \langle 0 0, +1 | 0 0, +1 \rangle}
               {  \langle J M, \Pi;p,h| J M, \Pi;p,h\rangle}
      \right]^\half \, .
\label{xi2}
\eeq
The two factors in Eq. (\ref{xi2}) are separately evaluated by
expanding both numerator and denominator in powers of the short-range
correlation function. The presence of the denominators is used to 
eliminate the unlinked diagrams \cite{fan87}.

Since in our calculations the correlation functions are purely scalar
they commute with the
operator $O_\eta (\bqu)$ and therefore we have to deal with terms of the
kind:
\begin{eqnarray}
\nonumber
\langle J M, \Pi;p,h|  O_{\eta} (\bqu) | 0 0, +1 \rangle &=&
\langle\Phi;J M, \Pi;p,h| F^\dagger 
O_{\eta} (\bqu) F |\Phi; 0 0, +1 \rangle_L \\
\nonumber
&=&  \langle\Phi;J M, \Pi;p,h| 
      O_{\eta} (\bqu) \prod^A_{i<j}f^2(r_{ij}) |\Phi; 0 0, +1 \rangle_L \\
&=&  \langle\Phi;J M, \Pi;p,h| 
O_{\eta} (\bqu) \prod^A_{i<j}(1+h_{ij})|\Phi; 0 0, +1 \rangle_L \, ,
\end{eqnarray}
where we have used the function $h_{ij}=f^2(r_{ij})-1$ and
the subindex $L$ indicates that only the linked diagrams are
considered. 

The approximation of our model consists in retaining only those terms
where the $h_{ij}$ function appears only once:
\begin{eqnarray}
\nonumber
\xi_{J,M,\eta,\Pi}(\bqu;p,h) & \longrightarrow &
\xi^1_{J,M,\eta,\Pi}(\bqu;p,h) \\
&\,=\,& \langle \Phi;J M, \Pi;p,h | \,O_{\eta}(\bqu)\, 
\sum_{i<j}\, (1+h_{ij}) \,|\Phi; 0 0, +1  \rangle_L \, .
\label{ximodel}
\end{eqnarray}
This result has been obtained using a procedure analogous to that
adopted in Ref. \cite{co95} for the evaluation of the density
distribution, and therefore the truncation of the expansion is done only
after the elimination of the unlinked diagrams.

We show in Fig. 2 the Mayer like diagrams describing
all the terms considered in our calculations.  This set of diagrams
has already been presented elsewhere \cite{co00,ama98}, but we show
them again because the identification of the individual diagrams is
essential for the discussion of the results.  In each diagram the
black square indicates the coordinate where the one-body operator
$O_{\eta}(\bqu)$ is acting, while the black dots indicate the other
coordinates.  The dashed line represents the correlation function
$h_{ij}$, which operates on two-coordinates only, and the continuous
oriented lines the single particle wave functions $\phi_k$.

\begin{figure}[t]
\begin{center}
\vspace*{1.5cm}
\hspace*{-1.0 cm}
\leavevmode
\epsfysize = 250pt
\epsfbox[70 200 500 650]{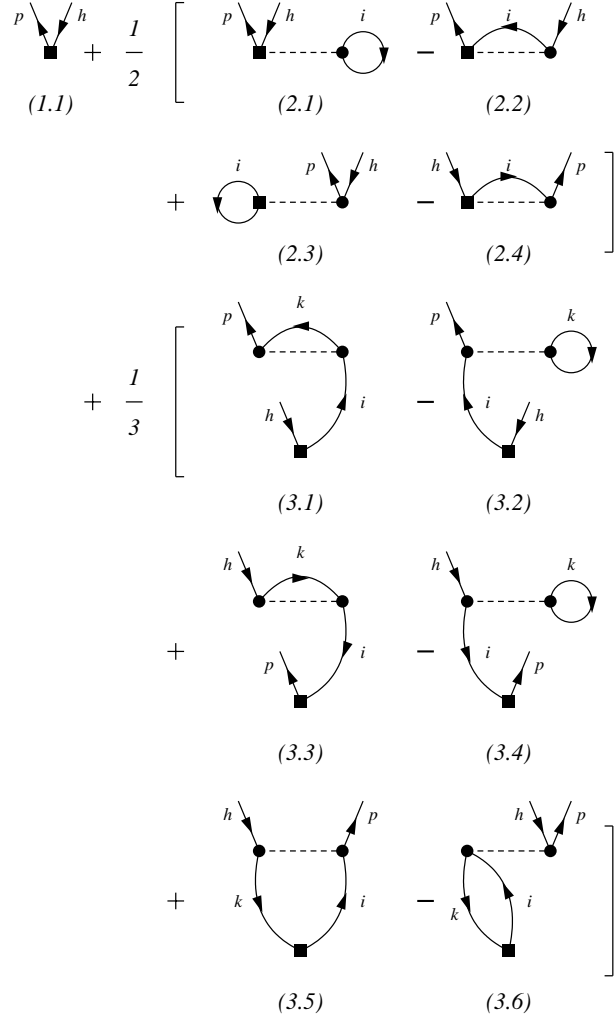}
\end{center}
\vspace{2.5cm}
\caption{\small Mayer-like diagrams considered. The black squares
represent the point where the external field is acting on, the dashed
line the correlation function and the oriented lines the single
particle wave functions.
}
\label{fig:diag1p1h}
\end{figure}

If we label with $1$ the coordinate where the external operator
$O(\bqu)$ is acting,  
we can specify the $\xi^1$ of Eq. (\ref{ximodel}) as:
\begin{eqnarray}
\nonumber
\xi^1_{J,M,\eta,\Pi}(\bqu;p,h) \, &=&\, 
\langle \Phi;J M, \Pi;p,h | \, O_{\eta}(\bqu)\,|\Phi; 0 0, +1  \rangle \\
\nonumber
\,&~& + \,  \langle \Phi;J M, \Pi;p,h |
        \, O_{\eta}(\bqu) \, \sum^{A}_{j>1}\, h_{1j}\, 
          |\Phi; 0 0, +1  \rangle_L     \\
\, &~& + \,  \langle \Phi;J M, \Pi;p,h |
\, O_{\eta}(\bqu) \, \sum^{A}_{1< i <j } \, h_{ij} \, 
 |\Phi; 0 0, +1  \rangle_L \, .
\label{xi1p1h}
\end{eqnarray}

The above expression shows that our model generates, 
in addition to the 
uncorrelated transitions represented in Fig. 2 
by the one-point diagram (1.1),
also two- and three-point diagrams. 
The presence of these last diagrams is
necessary to have the correct normalization of the many-body wave
function, as discussed in \cite{co00}.

In our case the operators $O_{\eta}(\bqu)$ are the Fourier transforms
of the charge and current operators $O_{\eta}(\br)$ given in 
Eqs. (\ref{charge1})-(\ref{magnetization1}). 
Since we describe
the nuclear transition between states with good
angular momentum, it is convenient to make a multipole expansions of
these operators. For the charge we have: 
\begin{equation}
M_{JM}(q) \,= \, \int {\rm d}^3r \, j_J(qr) \, Y_{JM}(\hat{\br}) 
 \, \rho(\br) \, ,
\label{mjm}
\end{equation}
where we have indicated with $\hat{\br}$ the $\theta$ and $\phi$
angles characterizing the vector $\br$ in polar coordinates and with
$j_J$ the spherical Bessel functions.

For the current we should distinguish between the electric excitations (E),
with natural parity $\Pi=(-1)^J$, and the magnetic excitations (M), with
unnatural  parity $\Pi=(-1)^{J+1}$:
\begin{equation}
T^{\rm E}_{JM}(q) \, = \, \frac{1}{q} \,
\int {\rm d}^3r \, \left\{ \nabla \times 
\left[ j_J(qr) \, {\bf Y}^M_{JJ}(\hat{\br}) \right ] \right\} 
\cdot {\bf J} (\br) 
\label{tejm}
\end{equation}
and
\begin{equation}
T^{\rm M}_{JM}(q) \, = \,
\int {\rm d}^3r \, j_J(qr) \,{\bf Y}^M_{JJ}(\hat{\br}) \cdot 
{\bf J} (\br) \, ,
\label{tmjm} 
\end{equation}
where we have used the symbol  ${\bf Y}^M_{JJ}$ to indicate the 
vector spherical harmonics \cite{edm57}.

By substituting the above equations into Eq. (\ref{xi1}) and  using
the Wigner-Eckart theorem and the properties of the 3j symbols, we
can rewrite Eq. (\ref{xi1}) in terms of reduced matrix elements: 
\begin{eqnarray}
\nonumber
\langle \Psi_f | O_{\eta} (\bqu) |\Psi_i \rangle &=& 
\frac{4 \pi}{|{\bf p}|} \sum_{p} \sum_{J,M,\Pi} {\cal A}(p,h;J,M,\Pi) \,\,
\xi^1_{J,M,\eta,\Pi}(\bqu;p,h) \\
\nonumber
&=& \frac{4 \pi}{|{\bf p}|} \sum_{p} \sum_{J,\Pi} {\cal A}
              (p,h;J,-\eta,\Pi) \,\, \\ 
\nonumber
&& \times (-i)^{-J} \sqrt{2 \pi} \biggl[ \langle \Phi; J, \Pi;p,h ||
\, O_{J(\eta)}(\bqu)\,|| \Phi; 0, +1 \rangle \\ 
\nonumber
\,&~& + \,  \langle \Phi; J, \Pi;p,h ||
        \, O_{J(\eta)}(\bqu) \, \sum^{A}_{j>1}\, h_{1j}\, 
          || \Phi; 0, +1 \rangle     \\
\, &~& + \,  \langle \Phi; J, \Pi;p,h ||
\, O_{J(\eta)}(\bqu) \, \sum^{A}_{1< i <j } \, h_{ij} \, 
 || \Phi; 0, +1 \rangle 
\biggr] \, ,
\label{xi11}
\end{eqnarray}
where we have defined the operators:
\begin{equation}
O_{J(\eta)}(\bqu)= \left\{ \begin{array}{ll} \sqrt{2} M_{J(0)}(q) &
    \textrm {if $\eta=0$}\\ 
T^{\rm E}_{J(\pm 1)}(q) + \eta T^{\rm M}_{J(\pm 1)}(q) & 
    \textrm {if $\eta=\pm 1$} \end{array} \right. \, .
\end{equation}
In Eq. (\ref{xi11}) the sum on M drops because $M=-\eta$ due to
angular momentum conservation.

Following Ref. \cite{co00}
the calculations of the transition matrix elements are carried out by
performing a multipole expansion of the correlation function $h_{ij}$:
\begin{equation}
h_{ij} \,= \,h(r_{ij}) \,= \, h(r_i,r_j,\cos \theta_{ij}) \,
 =  \, \sum_{L=0}^\infty  \, h_L(r_i,r_j)  \, P_L(\cos \theta_{ij}) \,
,
\label{hexp}
\end{equation}
where $P_L$ are the associated Legendre polynomials.

The final state
quantum numbers are determined by the quantum numbers
of the uncorrelated many body state $|\Phi^\Pi_{JM}\rangle$
built as a Slater determinant of single particle
wave functions of the form:
\begin{equation}
\phi_k(\br) \, \equiv \,
R^t_{n_k l_k j_k}(r) \, 
\sum_{\mu_k s} \, \langle  l_k \mu_k \half s | j_k m_k \rangle  \,
 Y_{l_k \mu_k}(\hat{\br}) \, \chi_s \, \chi_t \, ,
\label{spwf}
\end{equation} 
where $\chi_s$ and $\chi_t$ are
the spin and isospin wave functions respectively.

By using the orthonormality of the single particle basis, we rewrite
Eq. (\ref{xi11}) as: 
\begin{eqnarray}
\nonumber \langle \Psi_f | O_{\eta} (\bqu) |\Psi_i \rangle &=& \frac{4
\pi \sqrt{2 \pi} }{|{\bf p}|} \sum_{p} \sum_{J,\Pi} (-i)^{-J}{\cal
A}(p,h;J,-\eta,\Pi) \,\, \\ 
\nonumber && \times \biggl[ \langle p ||
\, O_{J(\eta)}(\bqu)\,|| h \rangle \, + \, \sum_{j}\, \langle p,j ||
\, O_{J(\eta)}(\bqu) \, h_{12}\, || h,j \rangle \\ 
\nonumber \, &~& +
\, \sum_{jk} \, \langle p,j,k || \, O_{J(\eta)}(\bqu) \, h_{23} \,
||h,j,k \rangle \biggr]\\ 
&=&\frac{4 \pi \sqrt{2 \pi} }{|{\bf p}|}
\sum_{ph} \sum_{J,\Pi} (-i)^{-J}{\cal A}(p,h;J,-\eta,\Pi) \,\,
\xi^1_{J,\eta,\Pi}(\bqu;p,h) \, ,
\label{xi111}
\end{eqnarray}
where $\xi^1_{J,\eta,\Pi}(\bqu;p,h)$ is the sum of the reduced
matrix elements corresponding to the one-, two- and three-points 
diagrams. Here the sums of
the $j$ and $k$ indices are running on all the hole single particle
wave functions.  
The subindices $1,2,3$ of the correlation function $h$ are associated
with the radial coordinates of the first wave function, where also 
$O_{J (\eta)}(\bqu)$ is acting, the second and the third one respectively.
We have indicated only the direct
matrix elements in the second and third term of the equation, but we
calculate all the exchange diagrams, as shown in
Fig. 2.

The multipoles of the one-body operators are
calculated by inserting the expressions (\ref{charge1}),
(\ref{convection1}) and (\ref{magnetization1}) into Eqs. (\ref{mjm}),
(\ref{tejm}) and (\ref{tmjm}). Generally, one may write the resulting
operator $O^{1}_{J M}(\bqu)$ as the product of a term depending on
$q$ and on the modulus of $\br_1$ times a term
depending only on the angular coordinates: 
\begin{equation}
{O}^{1}_{JM}(\bqu) \equiv F_J(qr_1){\cal O}_{JM}(\hat{\br}_1) \, .
\label{divop}
\end{equation}

The evaluation of the matrix elements of the various diagrams 
for the three operators follows the lines of Ref. \cite{co00}. 
In the present calculation we have added all the correlated diagrams 
of the convection current, which in that reference was included only
at the mean field level.

All the ingredients needed to calculate the cross section are now available.
The information on the angular distribution of the emitted particle is
fully contained in the spherical harmonics of Eq. (\ref{xixixi2}), which
has to be squared to calculate the response. We have to deal with a
geometrical term of the type:
\begin{eqnarray}
\nonumber
&&\hspace*{-2.5cm}
\biggl( (-i)^{-J'}{\cal A}(p',h';J',-\eta',\Pi') \biggr)^* \,\,
\biggl( (-i)^{-J}{\cal A}(p,h;J,-\eta,\Pi) \biggr) \, =   \\
\nonumber
&=&  
\langle \lp'  (-\mh -\eta' -\sigma) \half \sigma 
| \jp'  (-\mh -\eta') \rangle 
     \langle \jp' (-\mh -\eta') \jh \mh| J' -\eta' \rangle \\
\nonumber
&& \langle \lp  (-\mh -\eta-\sigma) \half \sigma | 
\jp  (-\mh -\eta) \rangle \langle \jp (-\mh -\eta) \jh \mh| 
J -\eta \rangle \\
&& (-i)^{\lp-\lp'-J+J'}  
Y_{\lp, (-\mh -\eta -\sigma) } (\what{\bpi})  
Y^*_{\lp',(-\mh -\eta' -\sigma) } (\what{\bpi}) \, .
\label{aapl}
\end{eqnarray}
This expression arises because the angular momentum of the hole
$\jh$ , its z axis projection $\mh$, and $\sigma$ the third component
of the spin of the emitted particle, 
are good quantum numbers of the
hadronic final state. The above equation shows that the dependence
from the angle $\phi_p$ between the scattering plane and the plane
where the momentum of the emitted particle is lying, is present only
when $\eta \neq \eta'$, i.e. only in the interference terms $tl$ and
$tt$ in Eq. (\ref{cross}).

We rewrite the cross section expression as:
\begin{eqnarray}
\nonumber
\frac{d^6 \sigma}{d \epsilon' d \Omega_e d E' d \Omega_p} &=&
\frac{\sigma_M K}{(2 \pi)^3} \left[
  v_l W_{0,0} + v_t (W_{+1,+1}+W_{-1,-1})  \right. \\
&~&
\left. + 2
v_{tl} Re(W_{0,-1}-W_{0,+1}) \cos \phi_p
+ 2 v_{tt} Re(W_{+1,-1}) \cos 2 \phi_p  \right],
\label{cross1}
\end{eqnarray}
where the $W$ functions are defined as \cite{co85}:
\begin{eqnarray}
\nonumber
W_{\eta \eta'} &=& \frac{32 \pi^3}{|{\bf p}|^2} \sum_{\sigma m_h}
\sum_{J \Pi \lp \jp} \sum_{J' \Pi' \lp' \jp'} (-i)^{\lp-\lp'-J+J'}
(-1)^{\eta-\eta'}  \\ 
\nonumber
&~& \langle \lp  (-\mh -\eta-\sigma) \half \sigma | \jp  (-\mh -\eta) \rangle 
     \langle \jp (-\mh -\eta) \jh \mh| J -\eta \rangle \\
\nonumber
&~& \langle \lp'  (-\mh -\eta' -\sigma) \half \sigma 
| \jp'  (-\mh -\eta') \rangle 
     \langle \jp' (-\mh -\eta') \jh \mh| J' -\eta' \rangle \\
\nonumber
&~& 
\left[ \frac {(\lp+\mh +\eta+\sigma)! } {(\lp-\mh -\eta-\sigma)! }
\frac {(\lp' +\mh +\eta' +\sigma)! } {(\lp' -\mh -\eta'-\sigma)!}
\right]^\half 
\left[ \frac{2 \lp+1}{4 \pi} \right]^\half 
\left[ \frac{2 \lp'+1}{4 \pi} \right]^\half \\
\nonumber
&~&
P^{-\mh -\eta -\sigma}_{\lp} (\cos \theta_p) \,\,
P^{-\mh -\eta' -\sigma}_{\lp'} (\cos \theta_p) \\
&~&
\xi^1_{J,\eta,\Pi}(\bqu;p,h) \, \,
\xi^{1 \dagger}_{J',\eta',\Pi'}(\bqu;p',h) \, .
\label{wpm}
\end{eqnarray}
\section{Specific applications}
\label{sa}
We have applied our model to the investigation of the
$^{16}$O(e,e'p)$^{15}$N reaction in the quasi-elastic region.  This
nucleus has been chosen for two main reasons.  The first one because
it is a sufficiently light nucleus, such that the plane wave
description of the electron wave functions is still valid
\cite{giu87}.  The second reason is related to the fact that this
nucleus has been studied with microscopic theories that provide us the
needed input for our calculations. Specifically, the single particle
bases and the correlation functions are taken from Ref. \cite{ari96}
and \cite{fab00} where they have been fixed by minimizing microscopic
many body hamiltonians. The calculations of Ref. \cite{ari96} have
been done with the semi-realistic nuclear interaction S3 of Afnan and
Tang \cite{afn68} by using a purely scalar correlation function.  The
calculations of Ref. \cite{fab00} have been done with a reduction of
the Argonne V18 interaction \cite{wir95} called V8' and used to
perform Green function Monte Carlo calculations \cite{pud97}. 
In \cite{fab00} a complicated state dependent correlation functions
was used, while in the calculations of the present paper we have
considered only the scalar term of this correlation. These two inputs
are the same used in Ref. \cite{co00} to calculate the inclusive
responses, and as in that case we shall label with S3 and V8 the
results obtained respectively with them.  For sake of completeness we
give in Table \ref{tab:pot} the parameters of the Woods-Saxon
potentials and we show in Fig. 3 the two correlation
functions used.  If not specifically mentioned our results have been
obtained by using the same mean field for particle and hole states. We
shall investigate in Sect. \ref{ort} the effect of using different mean
fields as it is commonly done in the investigation of (e,e'p) data.

\begin{figure}[hb]
\begin{center}
\vspace*{-3.5cm}
\hspace*{-.0 cm}
\leavevmode
\epsfysize = 300pt
\epsfbox[70 200 500 650]{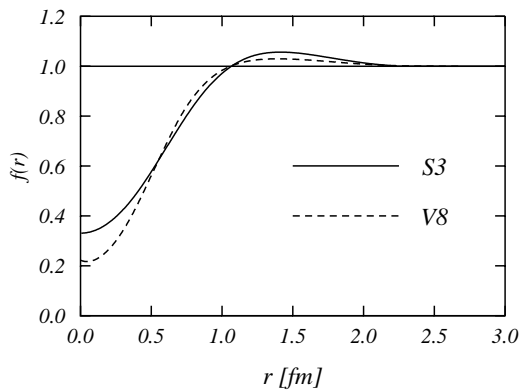}
\end{center}
\vspace{-2.2cm}
\caption{\small Correlation functions used in our calculations.
}
\label{fig:corr}
\end{figure}

\begin{table}[hb]
\caption{Parameters of the Woods-Saxon potentials used to generate the
  set of single particle wave functions.}
\label{tab:pot}
\begin{center}
\begin{tabular}{|c|l|rr|}
\hline
\multicolumn{2}{|c|}{ } &  V8   & S3 \\
\hline
 $\pi$ & $V_0$  [MeV]   & -53.0 & -52.5 \\
       & $V_{ls}$ [MeV] & 0.0   &  -7.0 \\
       & $R$   [fm]     &  3.45 &  3.2  \\
       & $a$   [fm]     &  0.7  &  0.53 \\
\hline
 $\nu$ & $V_0$  [MeV]   & -53.0 & -52.5  \\
       & $V_{ls}$ [MeV] & 0.0   &  -6.54 \\
       & $R$   [fm]     &  3.45 &  3.2   \\
       & $a$   [fm]     &  0.7  &  0.53  \\
\hline
\end{tabular}
\end{center}
\end{table}

Our calculations have been done in both perpendicular and parallel
kinematics. In the first situation the value of the momentum transfer
$\bqu$ is kept fixed and the changes of $p_i = |\bpi - \bqu|$, the
momentum of the proton before being emitted, are obtained by changing
the detection angle $\theta_p$. This is, for us, the easiest case to
calculate and to investigate, even though all the four response
functions $W_{\eta \eta'}$ contribute. In the second situation, the
parallel kinematics, the momentum transfer $\bqu$ and the momentum of
the emitted nucleon $\bpi$ are kept in the same direction and the
missing momentum is changed by changing the values of $\bqu$. In this
case the interference responses, those with $\eta \ne \eta'$ do not
contribute \cite{co85}.

Since in (e,e'p) processes
the number of variables into play is relatively
large, we have mainly done our investigation at a fixed value of 
$\omega= \epsilon_i - \epsilon_f$ = 128 MeV, and for a
fixed value of momentum of the emitted proton, $|\bpi|$= 444 MeV/c,
knocked out from the 1p$_{1/2}$ level. In perpendicular kinematics the
scattering angle was fixed to set $\bqu = \bpi$, to be able to scan
all the values of $p_i$ from 0 up to 2$\bqu$.
The kinematical conditions just described correspond, for the
perpendicular kinematics, to those of experimental data 
of Ref. \cite{ber82} and, for the parallel one, to the data of
\cite{leu94}. 

\begin{figure}[ht]
\begin{center}
\vspace*{-3.5cm}
\hspace*{-5.5 cm}
\leavevmode
\epsfysize = 280pt
\epsfbox[70 200 500 650]{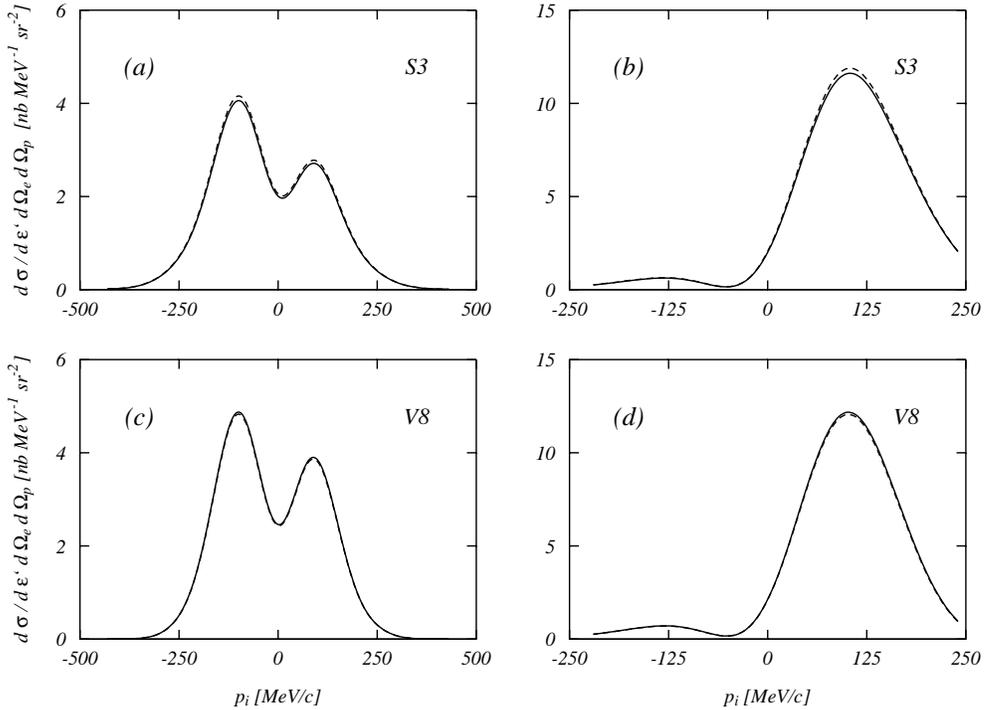}
\end{center}
\vspace{2.3cm}
\caption{\small Cross sections of the $^{16}$O(e,e'p)$^{15}$N process
for the kinematics described in the text as a function of the nucleon
initial momentum $p_i \equiv |\bpi_i|= |\bpi - \bqu|$.  The proton is
emitted from the 1p$_{1/2}$ level. The left panels have been obtained
in perpendicular kinematics while the other ones correspond to
parallel kinematics. The labels S3 and V8 refer to the two different
inputs as discussed in the text. The full lines represent the results
obtained with our model, while the dashed ones have been obtained
leaving out the three-body diagrams of Fig. 2. The mean-field results
are almost exactly overlapped to the full ones and are not included in
the figure.
}
\label{fig:cross}
\end{figure}

\subsection{General features of the results}
\label{gf}
In this section we discuss the general features of our results.  In
Fig. 4 we present the cross sections for the kinematics
described above as a function of $p_i$.  We should recall that we
defined $p_i \equiv |\bpi_i|=|\bpi - \bqu|$. In perpendicular kinematics,
we consider the values of $p_i$ to be positive when $\phi_p$=0 and
negative when $\phi_p$=180$^0$. In parallel kinematics $p_i>0$ when
$|\bpi| > |\bqu|$ and $p_i<0$ when $|\bpi| < |\bqu|$.  In Fig.
4, the left panels show the perpendicular kinematics
results while the right ones correspond to the parallel kinematics.
The full lines have been obtained considering the complete model.
Since the main novelty of our calculations is the presence of the
three-body diagrams, we shall discuss their role in some detail.
For this reason we show with dashed lines the results obtained by
adding to the mean field only those terms coming from the two-point
diagrams. The mean-field results we have obtained, are almost exactly 
overlapped to the full ones, and have not been shown.

The first remark we can make is that the effect of the correlations is
very small, as it was found in Refs. \cite{co00,mok00} for the
inclusive reactions. 
As already pointed out in Refs. \cite{co00,co95,ari97} the
three-point diagrams reduce the effect produced by the two-point
diagrams whatever it is. 
This is the consequence of the requirement of the
conservation of the number of particles. 

\begin{figure}[ht]
\begin{center}
\vspace*{-3.5cm}
\hspace*{-5.5 cm}
\leavevmode
\epsfysize = 280pt
\epsfbox[70 200 500 650]{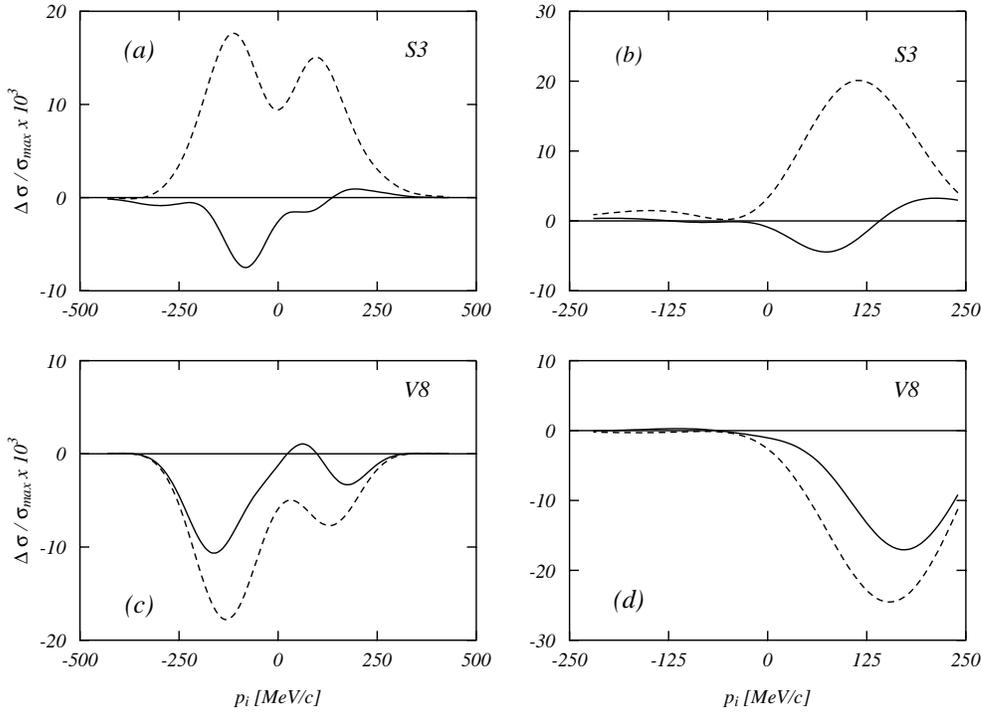}
\end{center}
\vspace{2.3cm}
\caption{\small Normalized differences between the correlated and
  uncorrelated cross sections of Fig. 4. 
 These are the differences between the cross sections 
 divided by the maximum values of the uncorrelated cross sections. 
 The dashed lines represent the
 two-point diagrams results while the full lines those obtained with
 the full calculations. The left panels show the perpendicular
 kinematics results and the right panels the parallel ones.
}
\label{fig:crossd}
\end{figure}

In order to emphasize the effect of the correlations we present in 
Fig. 5 the same results in terms of 
difference between the correlated cross sections and the uncorrelated
ones. In order to have a relative measure of the effect we have
divided these differences by the maximum values of the uncorrelated
cross sections. Henceforth we shall call these results normalized
differences.

The effect of the two-point diagrams is rather different for the S3
and V8 calculations. In the former case is positive, the mean field
cross section is increased by the correlations, while it is negative
in the latter case. It is interesting to notice that the final result
is rather similar in both cases as one would expect given the strong
similarity of the two correlation functions used.

The effect of the correlations on the individual responses calculated
in perpendicular kinematics is shown in Fig. 6. We should
first notice that the charge response $W_L \equiv W_{0,0}$ and the
transverse response $W_T \equiv W_{+1,+1} +  W_{-1,-1}$ are much
larger than the interference responses  
$W_{TL} \equiv 2 Re(W_{-1,0}- W_{0,+1})$
and  $W_{TT }  \equiv 2 Re (W_{+1,-1})$. In the charge
response  $W_L$ the various terms of the correlations behave like the
full cross sections, the final results being a lowering of the
mean-field response. This is not the case for the  $W_T$ response,
showing an increase of the mean-field results. 
This effect has been discussed in detail in Ref.\cite{co00} 
where it has been shown that it is
related to the absence of the diagram (2.3) of Fig. 2
in the longitudinal response.

In what refers to the interference responses, the behavior of
$W_{TL}$ is similar to that of $W_L$. In  $W_{TT }$ the correlations
increases the mean field responses. In this last responses the 3-point
diagrams appear to be very sensitive to the input, their contribution
is negligible for V8 calculation and noticeable for the S3 ones. 

\begin{figure}[ht]
\begin{center}
\vspace*{-3.5cm}
\hspace*{-5.5 cm}
\leavevmode
\epsfysize = 280pt
\epsfbox[70 200 500 650]{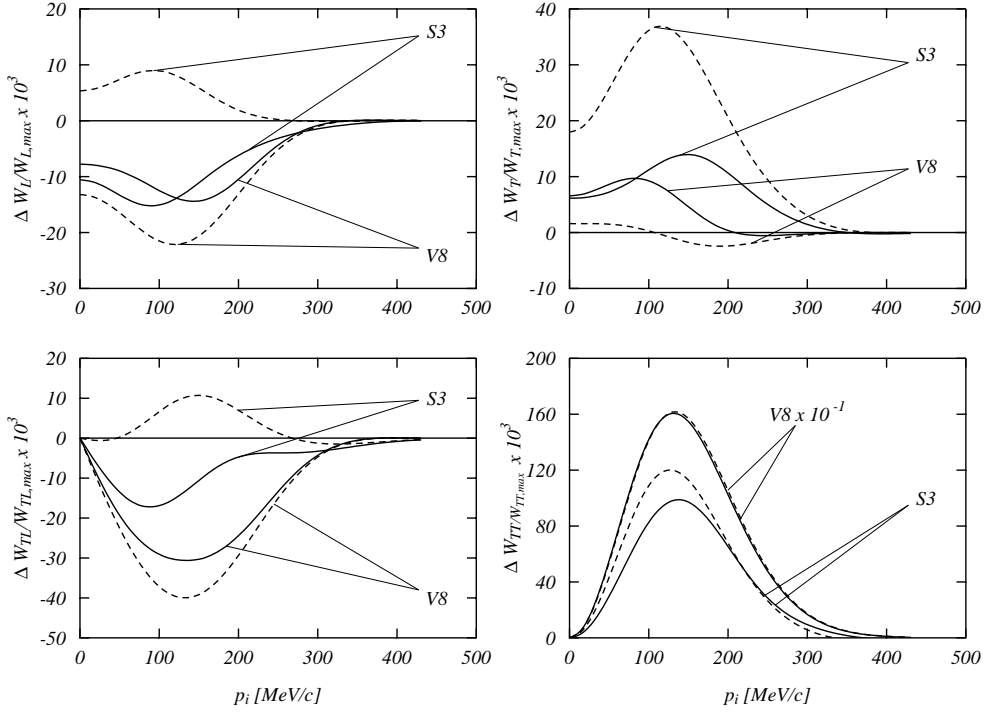}
\end{center}
\vspace{2.3cm}
\caption{\small Differences between correlated and uncorrelated
response functions in the perpendicular kinematics for
the calculations of Fig. 4. The differences have been
divided by the maximum value of the uncorrelated responses.
The full lines represent the results of the complete calculations  
and the dashed ones those obtained using the two-point diagrams only.
}
\label{fig:respd}
\end{figure}

\vspace*{5.5cm}

\begin{figure}[ht]
\begin{center}
\vspace*{-9.5cm}
\hspace*{-5.5 cm}
\leavevmode
\epsfysize = 280pt
\epsfbox[70 200 500 650]{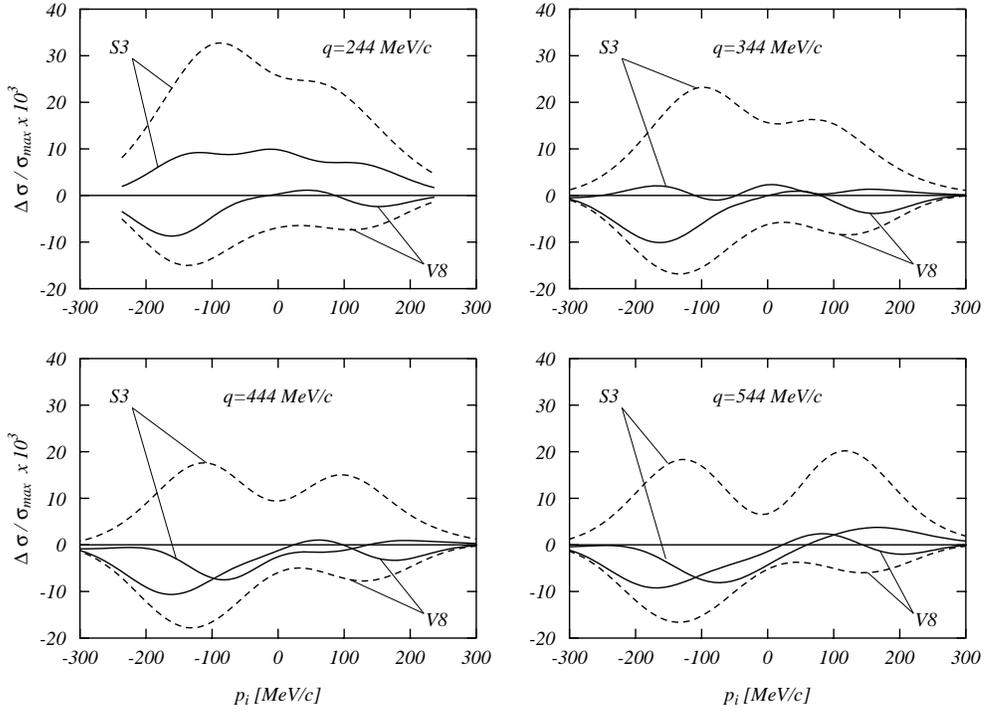}
\end{center}
\vspace{2.3cm}
\caption{\small Normalized differences between correlated and
uncorrelated cross sections calculated in perpendicular kinematics for
different values of the momentum transfer $q \equiv |\bqu|$.  As in
the previous figures the dashed lines refer to the two-point diagrams
calculations while the full lines show the results of the full
calculation.  
}
\label{fig:qdep}
\end{figure}

\pagebreak[4]

The cross section is obtained by summing the contributions of all the
responses.  
From Eq. (\ref{cross1}) one can observe that the $W_{TL}$ responses is
added for positive values of $p_i$ and subtracted for the negative
ones, and this generates the asymmetry in the perpendicular cross
sections of Fig. 4. 

We have investigated the momentum transfer dependence of the
correlation effects by calculating (e,e'p) cross sections in
perpendicular kinematics for different values of $|\bqu|$.  In all the
calculations we have fixed the momentum of the emitted nucleon such that
$|\bpi|=|\bqu|$. Results obtained for four different values of
$|\bqu|$ are shown in Fig. 7 as normalized differences.
As in the previous figures the dashed lines show the results
obtained by adding to the mean field calculation the two-point
diagrams only, while the full lines show the results of the full
calculation where also the three-point diagrams have been added.  The
shape of the lines is slightly modified at different values of
$|\bqu|$, but the order of magnitude of the effect is almost the same.
It is interesting to notice that the cancellation of the two-point
diagrams effects produced by the inclusion of the three-point terms
generates results rather independent from the correlation function.

\begin{figure}[ht]
\begin{center}
\vspace*{-3.5cm}
\hspace*{-5.5 cm}
\leavevmode
\epsfysize = 280pt
\epsfbox[70 200 500 650]{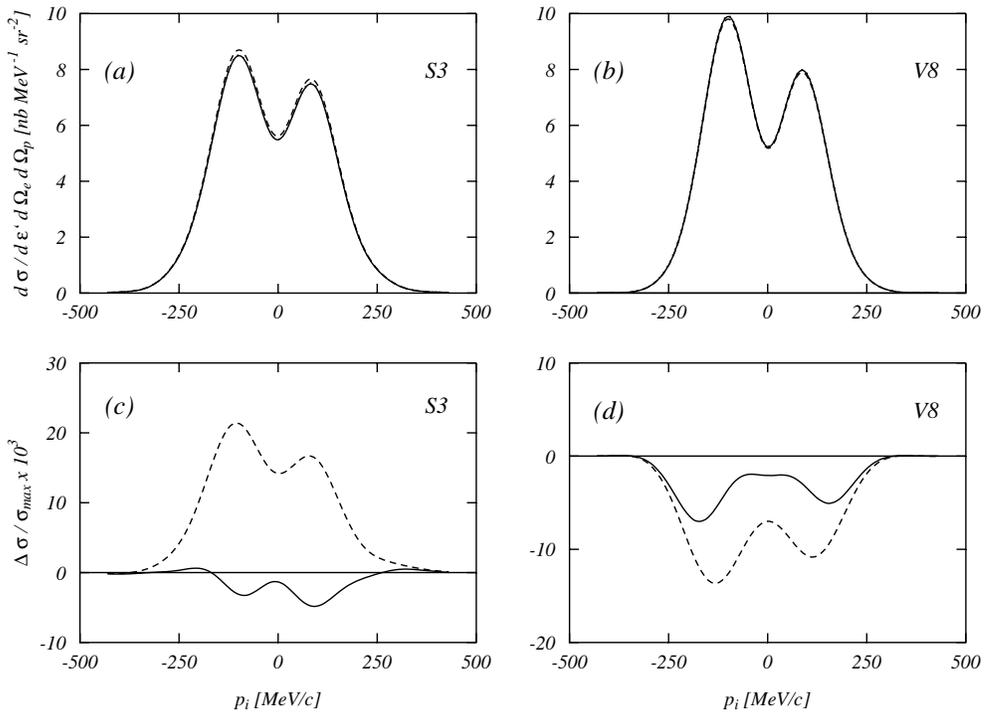}
\end{center}
\vspace{2.5cm}
\caption{\small
Cross sections for the emission of the 1p$_{3/2}$ proton
in perpendicular kinematics for the S3 and V8 input parameters.  The
dashed lines show the results obtained adding the two-point diagrams
to the uncorrelated terms and the full lines the result of the
complete calculation. The uncorrelated results are almost exactly
overlapped to the full ones and have not been shown. In the panels $a$
and $b$ we show the cross sections while in the panels $c$ and $d$ the
normalized differences between correlated and uncorrelated cross
sections.
}
\label{fig:p32}
\end{figure}

\begin{figure}[ht]
\begin{center}
\vspace*{-3.5cm}
\hspace*{-5.5 cm}
\leavevmode
\epsfysize = 280pt
\epsfbox[70 200 500 650]{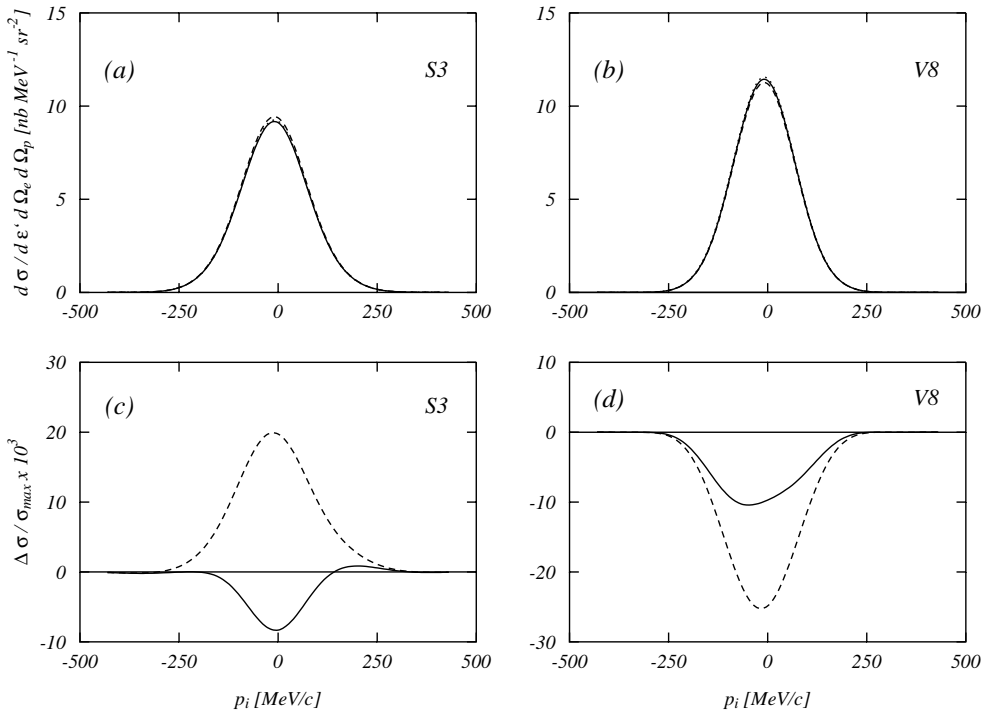}
\end{center}
\vspace{2.5cm}
\caption{\small The same as the previous figure for the emission of
  the 1s$_{1/2}$ proton.
}
\label{fig:s12}
\end{figure}

We have also verified the possible changes of the correlation effects
depending upon the hole wave function. As an example we show in Figs.
8 and 9 the cross sections, and the normalized
differences, obtained for the emission of the 1p$_{3/2}$ and
1s$_{1/2}$ protons under the same perpendicular kinematics of
Fig. 4. While for the 1p$_{3/2}$ case the shapes of the
cross sections are noticeably similar to those previously shown, the
curves obtained for the 1s$_{1/2}$ proton are rather different.  The
small differences in the radial shapes of the two p hole wave functions
do not modify sensitively the behavior of the cross section. On the
other hand, the big difference in the s wave cross section is due to
the different radial shape of the hole wave function. In spite of this
the behavior of the correlation effects is analogous in all the cases
we have considered.  The calculations done with two-point diagrams
show noticeable sensitivity to the change of the correlation, while,
when the three-point diagrams are added all the results are similar
and almost independent of the correlation function.

\subsection{Orthogonality and consistency of the input}
\label{ort}
In the analysis of the (e,e'p) experiments it is a common practice to
use a real mean field potential to describe the hole wave function
while the particle wave function is described using an optical
potential containing also an imaginary part.  This procedure, adopted
to take into account the final state interaction of the emitted
particle with the rest nucleus, produces a non orthogonality between
particle and hole single particle wave functions.  The effects of this
procedure on the (e,e'p) cross section have been thoroughly studied in
Ref.  \cite{bof82} where it has been shown that they are negligible
for the typical kinematics used in the experiments.

We have investigated the non orthogonality effects when correlations
are included in the calculations. In our derivation of the transition
matrix elements the orthogonality of the single particle wave
functions is essential. On the other hand, since we use central
potentials, only the radial parts of the wave functions change,
therefore all the angular momentum algebra is still valid. This means
that the non orthogonality is present only between single particle
wave functions
having the same angular momentum but different number of nodes in the
radial part.

We have done two calculations where the particle wave functions have
been evaluated with a different mean field with respect to the hole
wave functions. In the first one the particle mean field was set to
zero; therefore the emitted particle wave function was described as a
plane wave. This kind of calculation is commonly referred as the Plane Wave
Impulse Approximation (PWIA). In the second type of calculation we used the
optical potential of Ref. \cite{sch82} with the parameterization
adapted for $^{16}$O in Ref. \cite{leu94}.

\begin{figure}[ht]
\begin{center}
\vspace*{-3.5cm}
\hspace*{-5.5 cm}
\leavevmode
\epsfysize = 280pt
\epsfbox[70 200 500 650]{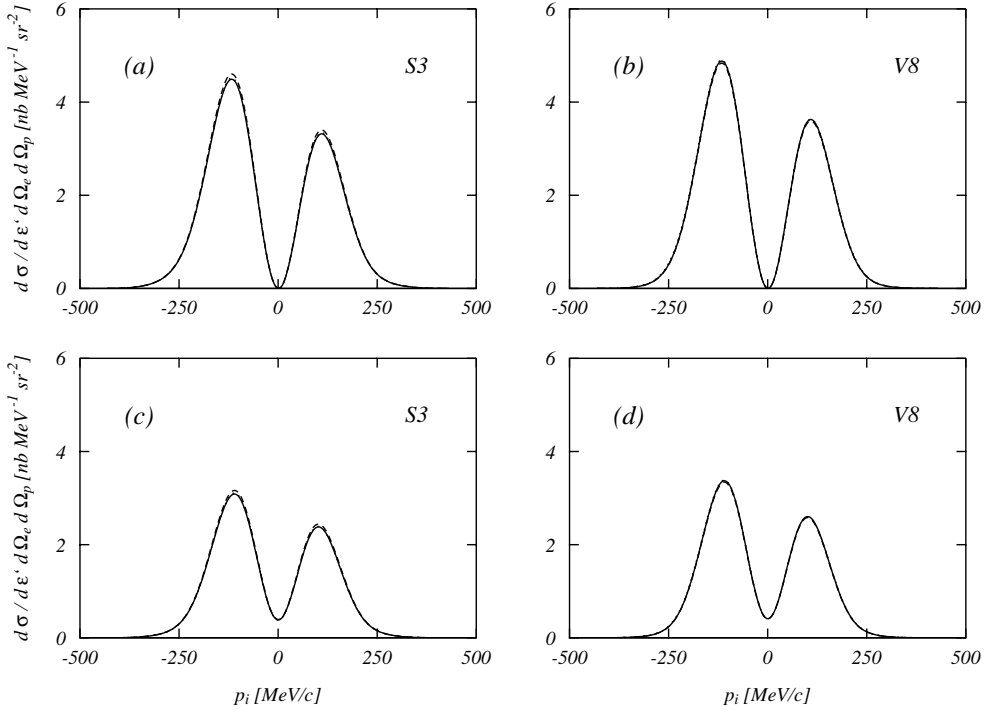}
\end{center}
\vspace{2.5cm}
\caption{\small Cross sections calculated in perpendicular kinematics
  under the same conditions of Fig. 4 with different
  mean field potentials for the particle wave functions. In panels $a$
  and $b$ the results obtained in Plane Wave Impulse Approximations
  are presented.  In panels $c$ and $d$ those obtained the optical
  potential of Schwandt et al. \cite{sch82}.  The meaning of the line
  is the same as in Fig. 4.
}
\label{fig:xopt}
\end{figure}

We show in Fig. 10 the cross sections calculated with the
S3 and V8 inputs for the same perpendicular kinematics conditions used
in Fig. 4.  The panels $a$ and $b$ show the PWIA
results, while the other two panels those obtained with the optical
potential. A comparison with Fig. 4 shows that the PWIA
cross sections are larger than the original ones, while the cross
sections obtained with the optical potential are smaller. It is
interesting to notice that the PWIA results have a minimum for
$p_i$=0 which is exactly zero, at expected, because in this approximation
the cross section is
related to the Fourier transform of the hole wave function.

\begin{figure}
\begin{center}
\vspace*{-3.5cm}
\hspace*{-5.5 cm}
\leavevmode
\epsfysize = 280pt
\epsfbox[70 200 500 650]{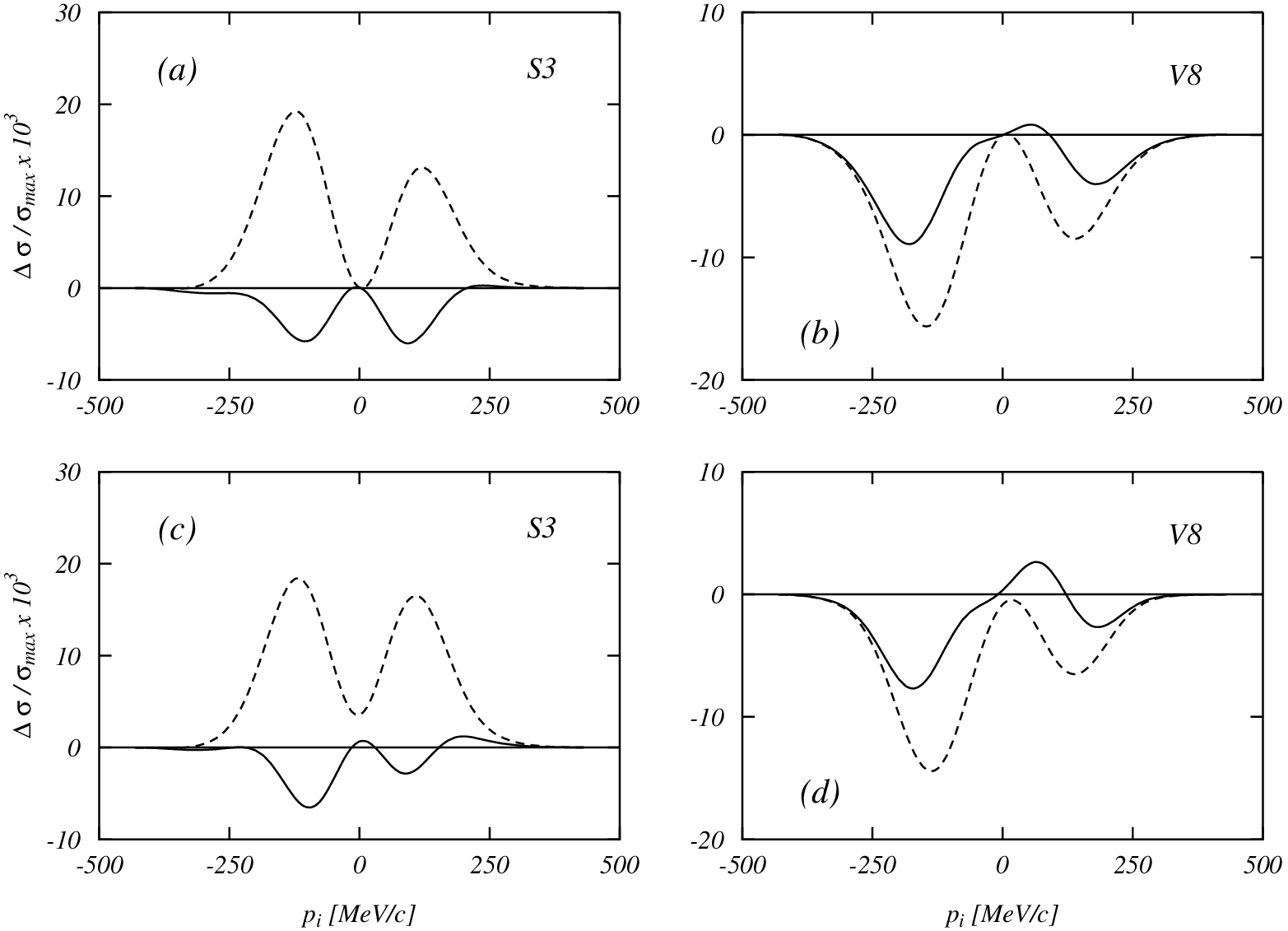}
\end{center}
\vspace{2.5cm}
\caption{\small Normalized differences between the cross sections
  presented in Fig. 10. The meaning of the lines is the
  same as in Fig. 5. 
}
\label{fig:dopt}
\end{figure}

\vspace{5.5cm}

\begin{figure}
\begin{center}
\vspace*{-3.5cm}
\hspace*{-5.5 cm}
\leavevmode
\epsfysize = 280pt
\epsfbox[70 200 500 650]{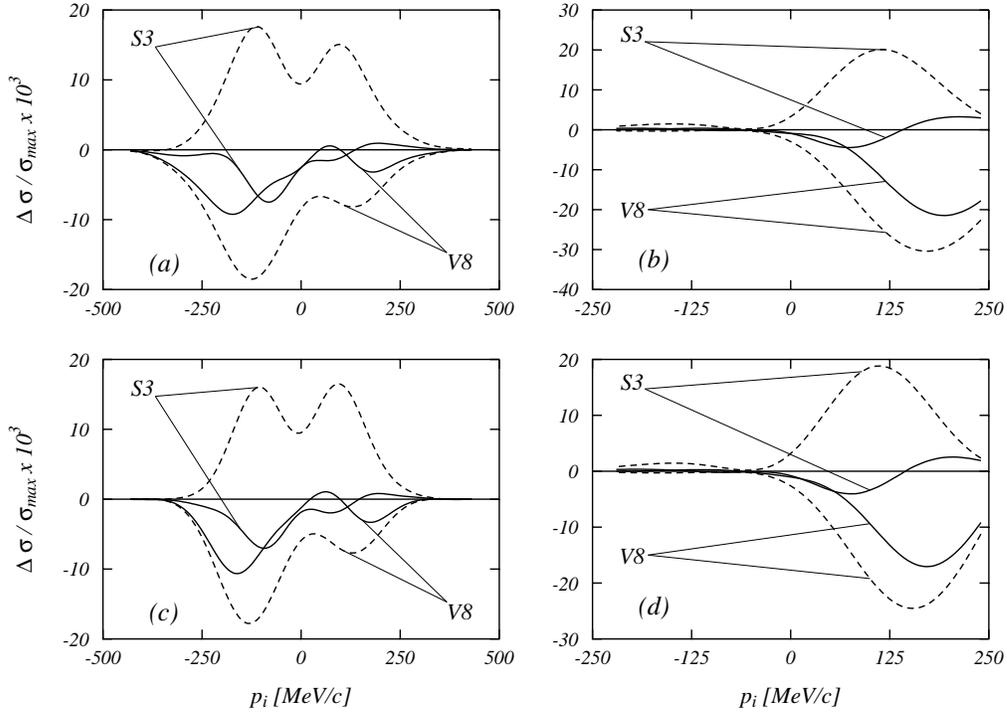}
\end{center}
\vspace{2.5cm}
\caption{\small Normalized differences between cross sections
  calculated with the same mean field potential but with different
  correlation functions.  In panels $a$ and $b$ the S3 mean field
  has been used, while in panels $c$ and $d$ the V8 mean field was
  considered.  The S3 and V8 labels indicate the correlation function
  used.  The calculations of the $a$ and $c$ panels have been done in
  perpendicular kinematics and those of the $b$ and $d$ panels in
  parallel kinematics. As usual the dashed lines refer to calculations
  with two-point diagrams while the full lines represent the complete
  calculations.
}
\label{fig:mix}
\end{figure}

In both kinds of calculations the effect of the correlations is very
small and to emphasize it we show in Fig. 11 the
normalized differences.  The behavior of the various terms is similar
to that shown in Fig. 5.  The results obtained with the
two-point terms are rather different when calculated with S3 and V8
inputs. When also the three-point diagrams are included they become
similar. Clearly the effects of the non-orthogonality of the basis are
negligible on the correlations.  We have obtained similar results also
in parallel kinematics.

A comparison between the three calculations done in perpendicular
kinematics shows that the shapes of the results obtained in PWIA and
with the optical potential are rather similar and quite different from
the results obtained with the same real potential used for the hole
states. This arises because the real
part of the Scwhandt potential is quite weak, $-28.359$ MeV, against the 
about $-50$ MeV of the potential used in our V8 and S3 calculations. The
effect of the optical potential is to conserve the PWIA shapes but to
lower the values of the cross section. This second effect is obtained
because of the presence of the imaginary part of the potential.

We have already stressed that our inputs are taken from microscopic
FHNC calculations where, for a given hamiltonian,
the energy mean value has been minimized by varying both mean field
and correlation function. This consistency of the input is not usually
respected in (e,e'p) calculations with short-range correlations, where
mean field and correlation functions are taken from different sources.

We have tested the need of relating mean-field and correlation
functions by interchanging them in our inputs. This means that we have
done calculations using the S3 mean field potential together with the
V8 correlation functions, and vice versa. The results of these
calculations are shown in Fig. 12 as normalized differences
and are compared in each panel with the results of Fig.
5.  Also in this figure the need of including
three-point diagrams for the reasons already mentioned becomes
evident. The final results seems to be rather independent of the
mean field used.  On the other hand, the quantity shown has been
defined to minimize the effects of the mean field and enhance those of
the correlations.

\subsection{Comparison with other approaches}
\label{coa}
In this section we shall try to make a comparison between our approach
and other models that handle short-range correlations in electron
scattering.  Depending upon the methodology used to attack the problem
we classify these models into two categories.  A first kind of
calculations \cite{jan00} uses an approach similar to ours but the
evaluation of the diagrams remains restricted to the so-called single
pair approximation. This approximation implies that only those
two-point diagrams where the knocked out particle is directly
connected to the electromagnetic operators are considered. In our
classification scheme these are the diagrams (2.1) and (2.2) of Fig.
2, even though the exchange diagram (2.2) is not
always considered.

The second category of calculations \cite{ami97,gai00} is
based upon the idea of the spectral function
\cite{fru84,bof96}. The basic hypothesis is that the nuclear
finite state wave function can be separated in terms of emitted
particle wave function and spectral function which is related to the
overlap between the wave function of the target nucleus and that of
the A-1 nucleon system.  Spectral functions used in the evaluation of
(e,e'p), and also (e,e'2N), cross sections have been calculated in
various manners.  In some case the Brueckner G-matrix theory has been
used \cite{ami97}, in others they have been obtained from 
variational Monte Carlo calculations \cite{lap99}. In the work of Ref.
\cite{gai00} the spectral functions have been obtained using the
asymptotic properties of the one-body density matrices obtained from
microscopic theories. A precise comparison with our approach is rather
difficult since the spectral functions include high order correlation
terms which we do not consider. What remains, however, is the fact
that also this approach considers only those diagrams where the emitted
particle is directly connected to the electromagnetic operators. In
our approach this could be translated by saying that in addition to
the diagrams (2.1) and (2.2), already mentioned, also the diagrams 
(3.3) and (3.4) are considered.

We have tried to mimic the other approaches by repeating the
calculations of Fig. 4 without those diagrams where the
particle is not directly connected to the electromagnetic operators.
In these calculations the 1/2 and 1/3 terms multiplying the two and
three point diagrams respectively have been set to 1, as is done in
the literature.

\begin{figure}[ht]
\begin{center}
\vspace*{-3.5cm}
\hspace*{-5.5 cm}
\leavevmode
\epsfysize = 280pt
\epsfbox[70 200 500 650]{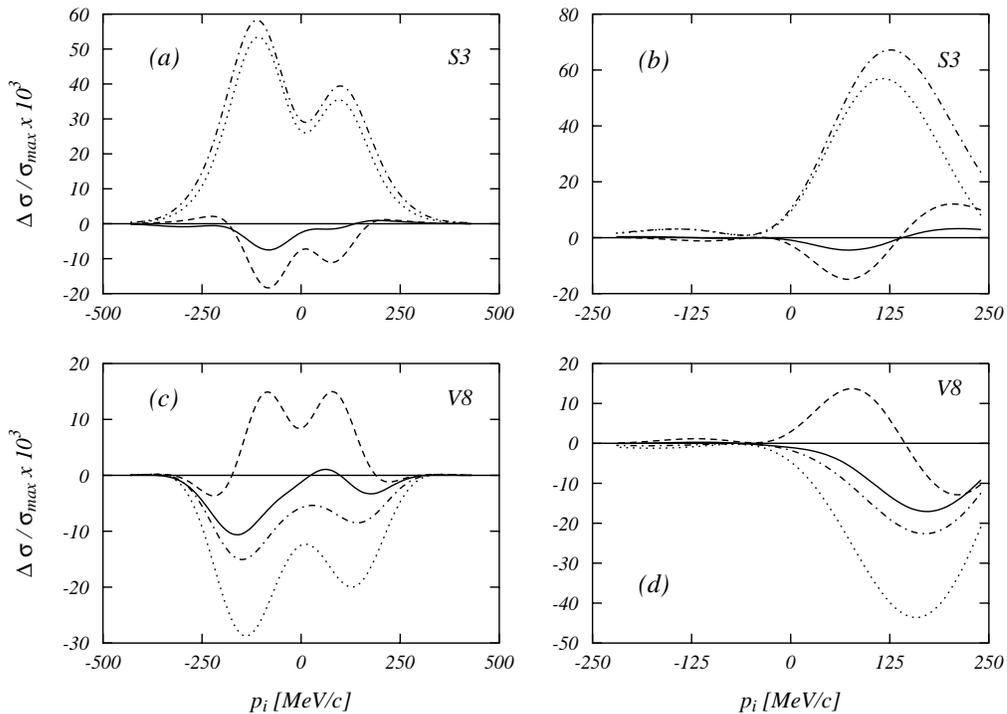}
\end{center}
\vspace{2.5cm}
\caption{\small Normalized differences between correlated and
  uncorrelated cross sections. The full lines represents the complete
  calculations, the dotted lines have been evaluated by adding to the
  mean field the contribution of the diagram (2.1) of
  Fig. 2, the dashed dotted lines, the contribution of
  the diagrams (2.1) and (2.2) and the dashed lines, the diagrams
  (2.1), (2.2), (3.4) and (3.5).
}
\label{fig:chdia}
\end{figure}

The results of these calculations are presented in Fig.
13 as normalized differences. The
full lines show the results of the complete calculations, those of
Fig. 4. The dotted lines have been obtained by using the
diagram (2.1) only, the dashed dotted lines have been obtained by
adding also the diagram (2.2) and the dashed lines by adding the 
diagrams (3.3) and (3.4).

The inclusion of the diagrams (3.3) and (3.4) changes radically the
effect of the correlations, modifying shapes and signs.  The final
result in the case of S3 is rather similar to that of the complete
calculation, but for the V8 input there is a noticeable difference.
This instability of these results shows, in a clear manner, the need 
of including all the terms of the expansion.
\subsection{Comparison with experimental data}
\label{ced}
In this section we compare the results of our calculations with some
experimental data available on $^{16}$O. In the literature
these data are presented in terms of a reduced cross section obtained by
dividing the cross section by the electron-nucleon cross section and
the kinematic factor $K$ of Eq. (\ref{cross}). In our calculations
we have used the electron-nucleon $\sigma_{cc1}$ cross section of
Ref. \cite{for83}.

In Fig. 14 we make the comparison with the 1p$_{1/2}$ proton
emission data of Ref. \cite{ber82} measured in perpendicular kinematics.
The full lines represent the complete calculations
of Fig. 4, where all the diagrams are considered and the
same mean-field potential for both particles and hole has been used.
In the panels $a$ and $b$ the results for the S3 and V8 inputs are
shown respectively. In the panels $c$ and $d$ the same curves have
been multiplied by reduction factors to improve the agreement with
the experimental data. These factors are 0.58 for the S3 result and
0.5 for the V8 one. Even after quenching the curves the agreement
with the data is rather poor. 

\begin{figure}[ht]
\begin{center}
\vspace*{-3.5cm}
\hspace*{-5.5 cm}
\leavevmode
\epsfysize = 280pt
\epsfbox[70 200 500 650]{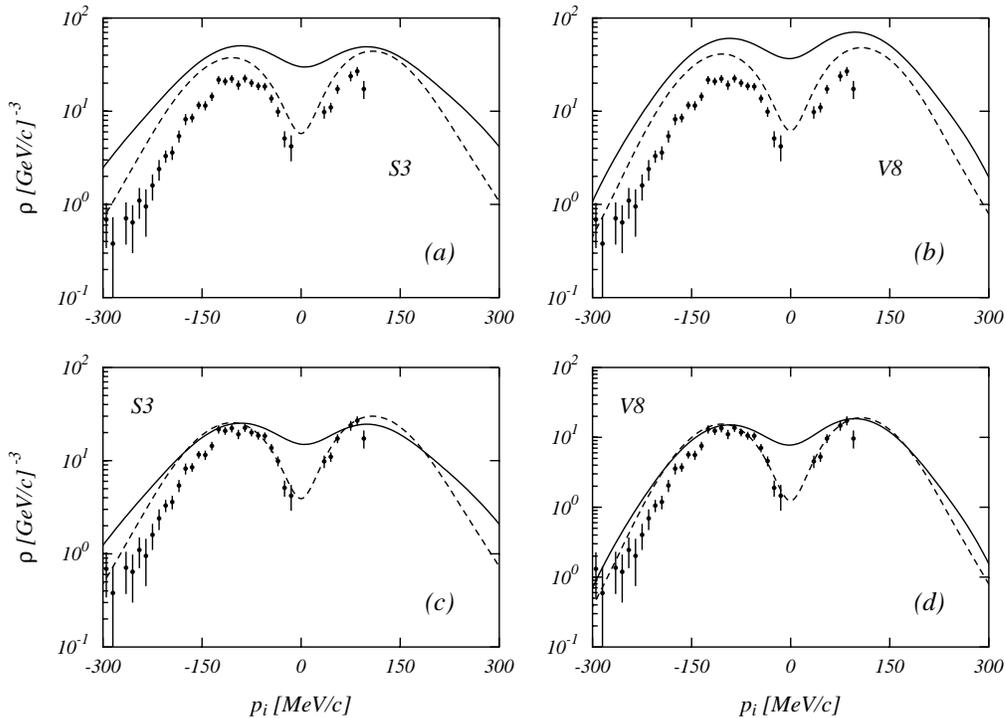}
\end{center}
\vspace{2.5cm}
\caption{\small Reduced cross section compared with the experimental
  data of  Saclay \cite{ber82}. The full lines have been obtained from
  the complete calculations of Fig. 4 in perpendicular
  kinematics. The dashed lines have been calculated with the optical
  potential of Ref. \cite{sch82}. In
  panels $c$ and $d$ the curves of the upper panels have been
  multiplied by reduction factors to reproduce the data (see text). 
}
\label{fig:sac1}
\end{figure}

The dashed lines of the figure have been obtained by changing the
particle mean field with the optical potential of Schwand et
al. \cite{sch82}, in the parameterization of Ref. \cite{leu94}. 
Also in this case a quenching of the curves is necessary, and the
final results, shown in the two lower panels, have been obtained
multiplying by 0.58 and 0.5 the S3 and V8 results respectively. The
dashed curves show a better agreement with the experimental data. 

\begin{figure}[ht]
\begin{center}
\vspace*{-3.5cm}
\hspace*{-5.5 cm}
\leavevmode
\epsfysize = 280pt
\epsfbox[70 200 500 650]{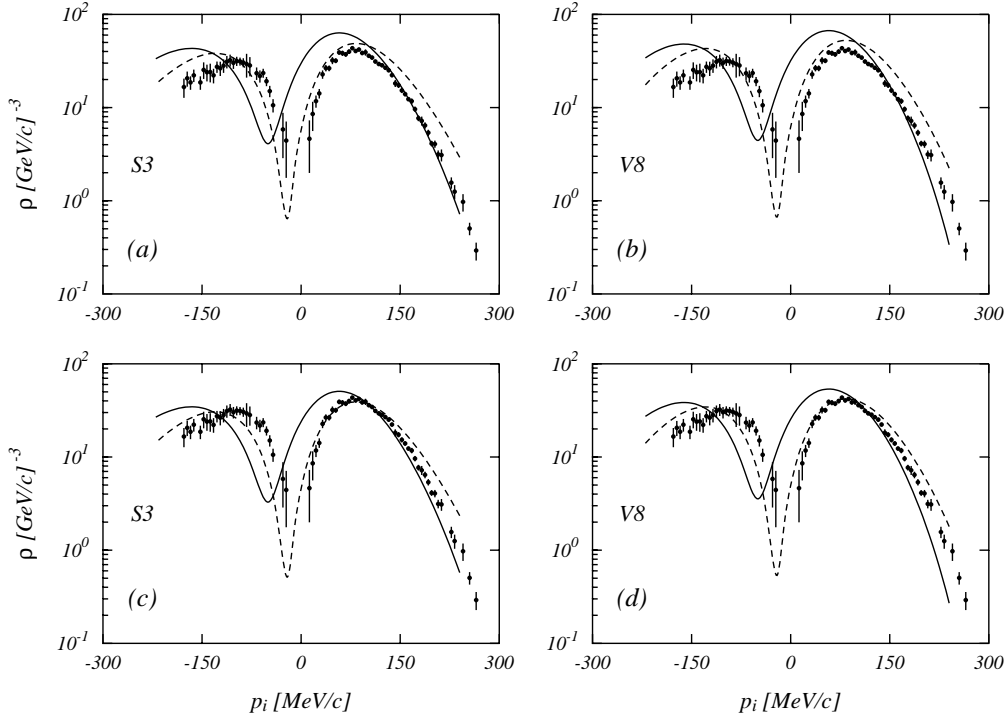}
\end{center}
\vspace{2.5cm}
\caption{\small Reduced cross sections compared with the experimental
  data of Nikhef \cite{leu94}. The full lines have been obtained from
  the complete calculations of Fig. 4 in parallel
  kinematics. The dashed lines have been calculated with the optical
  potential of Ref. \cite{sch82}. In
  panels $c$ and $d$ the curves of the upper panels have been
  multiplied by a common reduction factors of 0.8. 
}
\label{fig:nik1}
\end{figure}

An analogous situation is found in Fig. 15 where the
comparison is done with the data of Ref. \cite{leu94}, again for the
emission of the 1p$_{1/2}$ proton but measured in parallel kinematics.
The meaning of the curves is the same of the previous figure, and in
this case we used for all the curves the same quenching factor of 0.8.

The results presented in Figs. 14 and 15
clearly show that the fully consistent calculations are unable to
reproduce the data. Not only in terms of their magnitude, but the obtained
results are qualitative different with respect to the data
distribution. 
The mean fields fixed by the minimization process in Ref. \cite{ari96}
and \cite{fab00} are too strong to describe properly the motion of the
emitted particle in the continuum as we have already observed.
The shift of the minimum of the reduced response with respect to $p_i=0$
in parallel kinematics is an effect produced by the distortion
potential. 

\begin{figure}[ht]
\begin{center}
\vspace*{1.cm}
\hspace*{-1.0 cm}
\leavevmode
\epsfysize = 280pt
\epsfbox[70 200 500 650]{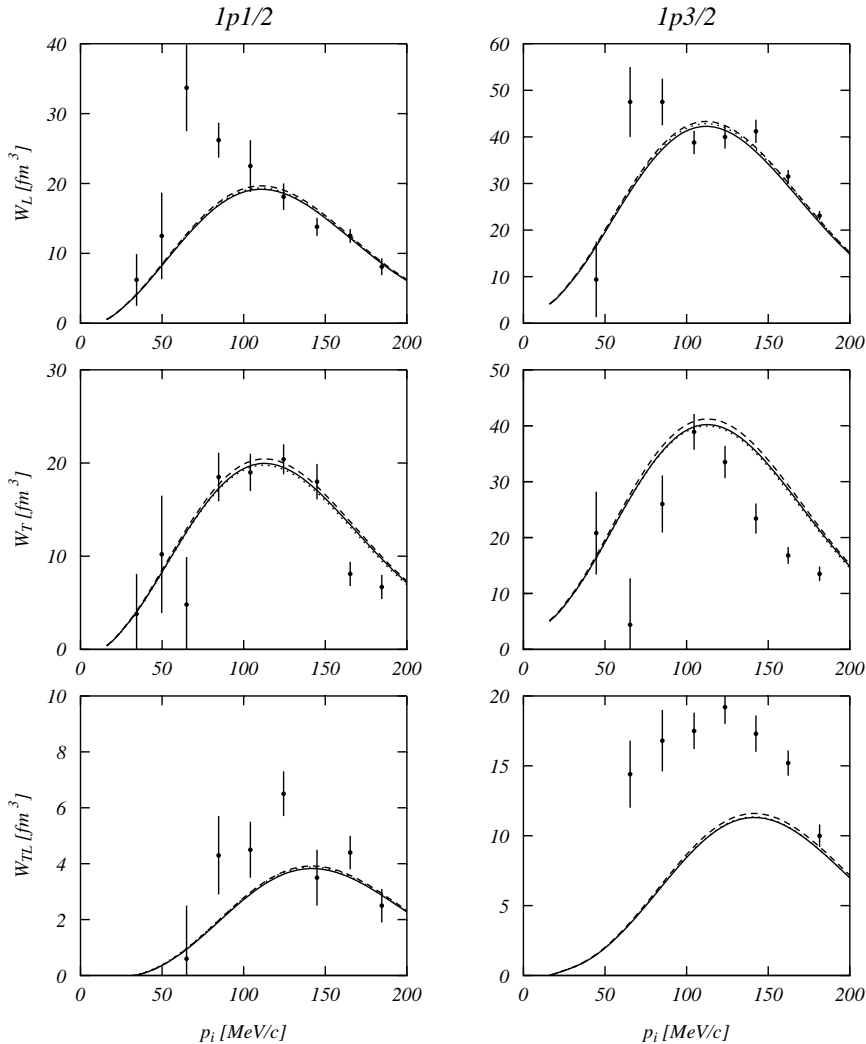}
\end{center}
\vspace{2.cm}
\caption{\small Response functions compared with the experimental data
  of Ref. \cite{spa93}. Our calculations have been done with the S3
  input and using the optical potential of Ref. \cite{sch82}. The full
  lines show the results of the complete calculations, the dashed ones
  those obtained with the two-point diagrams, and the dotted lines
  (overlapped to the full ones), the mean-filed results. The
  panel labelled $W_{L}$ shows actually 
  $W_{L} + \half \frac {\bqu^2} {\bqu^2 - \omega^2} W_{TT}$. The
  contribution of $W_{TT}$ is in any case small. The response $W_{TL}$
  is multiplied by $1/\sqrt{2}$ in order to agree with the response
  experimentally separated. All the curves have
  been multiplied by a 0.7 factor.
}
\label{fig:wexp}
\end{figure}

In Fig. 16 we compare the results of our calculations with the
empirical values of the electromagnetic responses given in
Ref. \cite{spa93}. These calculations have been done with the S3 input
and using the Scwhandt et al. optical potential \cite{sch82}. The
meaning of the various lines is that of Fig. 4: full lines show the
complete calculations and dashed lines have been obtained by adding
the two-points diagrams to the mean-field terms. We do not draw the
mean-field results because they are almost overlapped with the full
ones. All the lines have been multiplied by a reduction factor of
0.7. The effect of the short range correlations is very small. One can
clearly identify the increasing of the responses produced by the
inclusion of only the two-point diagrams, but the final results are
essentially back to mean-field result.  The calculation done with the
V8 input produces analogous results, with the obvious difference in the
two-point calculations.  The same data have been studied in
Ref. \cite{ama99} where it has been shown that the effects of the
Meson Exchange Current are larger than those we find for the
short-range correlations.
\section{Summary and conclusions}
\label{sum}
In this work we have studied the effects of short-range correlations
on the one-nucleon emission processes induced by electromagnetic
probes. The description of these processes has been done by using a
nuclear model that considers all the linked diagrams containing a
single correlation line. This implies that in addition to the
two-point diagrams, commonly considered in the literature, also
three-point diagrams should be included in order to maintain the
proper normalization of the many-body wave function.

The calculations have been done using single particle wave functions
and correlations taken from microscopic FHNC
calculations \cite{ari96,fab00}. We have considered only the
scalar term of the correlation. We have calculated the  
$^{16}$O (e,e'p) $^{15}$N reaction for the perpendicular and parallel
kinematics of the Saclay and Nikhef experiments of Refs. \cite{ber82}
and \cite{leu94}.

In the kinematic conditions investigated, the effects of the
short-range correlations are extremely small as compared to the
mean-field results. They are within the accuracy of the experimental
data and the uncertainty in the theoretical inputs.

The two correlation functions used in our calculations are rather
similar, and therefore we expect them to produce similar effects on the
mean--field results. This does not happens if only two-point diagrams
are included. The S3 increases the mean-field responses, and cross
sections, while the V8 lowers their values. The effects of the
correlations become similar when the three-point diagrams are
included. The detailed study of the separated responses shows that the
three-point diagrams reduce the effect produced by the two-point
diagrams alone, whatever they are. This fact is a consequence of the
need of the three-point diagrams to ensure the proper normalization of
the wave function. This result is almost independent of the
momentum transfer or hole wave function. 

We have been concerned about the consistency of the input, and we have
calculated cross sections with different mean fields describing the
emitted particle, and we have also interchanged correlation functions
and mean field in our inputs.  We did not find significant changes on
the effects of the correlations, saying that these effects are rather
independent of the chosen mean field.

We have tried to compare our short-range correlation model with other
approaches one finds in the literature. In these approaches only
those diagrams where the emitted particle is directly connected to the
one-body electromagnetic operator are considered. We have done
calculations where only these diagrams have been considered, and we
found a strong dependence of the results on the number of diagrams
included, showing big instabilities to small changes of the input.
The situation is rather complicated by the fact that in these
approaches the particle wave function is usually described as moving
in an optical potential. Therefore it is not clear whether the
diagrams we have calculated explicitly are effectively considered by the
optical potential.

The comparison with the experimental data of Refs. 
\cite{ber82,leu94,spa93}, shows that the short-range correlation
effects do not significantly improve the agreement with the data. In
this respect various problems remain open, such as the physical origin
of the required quenching or the reason why the parameters of the
optical potential are so different from those describing the hole
states. The investigation of these problems is beyond the scope of the
present article.

Finally we would like to discuss some of the limitations of our
calculations related to the hypotheses and the approximations of
our model.  One of these limitations is due to the fact that we
consider only those diagrams containing a single correlation function.
Clearly higher order terms could be important. We have already
mentioned that the test of the validity of our model was done in Ref.
\cite{ama98} for the inclusive nuclear matter charge response. The
excellent agreement between our results and those obtained considering
an infinite set of diagrams, give us confidence about our
approximation, and we trust its validity also in the case of finite
nuclear systems and for the one-body current operators. It would be
interesting, in any case, to make other tests on it.

Another limitation of our approach is related to the use of
correlation functions of purely Jastrow type.  Recent calculations
show that the tensor correlations on two-nucleon emission processes
\cite{jan00,giu99} produce effect comparable, or even larger, than
those induced by the scalar correlations.  However, these calculations
have been done in the single pair approximation, and we have shown how
dangerous is to rely on such an approximation which can produce large
effects since the three-point counter terms are not present.  On the
other hand, we know from microscopic calculations \cite{fab98} that
the tensor correlations are essential to bind nuclei. They are also
important, together with Meson Exchange Currents, in the nuclear
matter transverse quasi-elastic response \cite{fab97} and they are
necessary to obtain a proper description of the electromagnetic
responses of few-body systems \cite{car98}. The possible effect of the
tensor correlations on nuclear responses is an open problem which
should be investigated.

Last, but not least, our model does not take into account those
processes beyond the mean-field description of the nucleus falling
under the generic name of long-range correlations. These correlations
are related to the coupling between the single particle dynamics and
the collective excitation modes of the nucleus. Their effects on
(e,e'p) processes have been considered in a nuclear matter approach
\cite{ami97}, but they do not seem to be relevant. However, one should
consider that collective surface vibrations are not present in nuclear
matter calculations, and they can be extremely important in finite
nuclear systems, as an investigation on the charge density
distribution differences of heavy isotones has recently
pointed out \cite{ang01}.

\vskip 2.0 cm
{\bf Acknowledgments}\\
We thank C. Giusti for useful discussions and M. Leuschner
for sending us the experimental data.  One of us, S.R.M.,
thanks the Italian Foreigner Ministry for the financial support
during his stay in Lecce. This work has been partially supported by
the agreement INFN-CICYT, by the DGES (PB98-1367), by the Junta de
Andaluc\'{\i}a (FQM225) and by the MURST trough the {\sl PRIN:
Fisica teorica del nucleo atomico e dei sistemi a molticorpi}.

\vspace*{8cm}


\end{document}